\documentclass[a4paper]{jpconf}
\usepackage{graphicx}
\usepackage{amsmath}
\usepackage{units}
\usepackage{feynmp}
\usepackage{epstopdf}
\usepackage{ifpdf}
\usepackage{braket}

  \ifpdf
  \fi

\newcommand{\be}{\begin{equation}}
\newcommand{\ee}{\end{equation}}

\def\beq{\begin{equation}}

\def\eeq{\end{equation}}

\def\bea{\begin{eqnarray}}

\def\eea{\end{eqnarray}}
\newcommand{\newc}{\newcommand}

\newc{\ifb}{\textrm{fb}^{-1}}

\newc{\bbdecay}{0\nu\beta\beta}

\newc{\betadecay}{0\nu\beta\beta}

\newc{\bhalflife}{T^{0\nu\beta\beta}_{1/2}(\Ge)}

\newc{\mee}{m_{\beta\beta}}

\newc{\Ge}{^{76}\textrm{Ge}}

\newc{\bbbar}{B^0_d\textrm{-}\bar{B}^0_d}

\newc{\kkbar}{K^0\textrm{-}\bar{K^0}}

\newc{\solar}{\Delta m^2_{\odot}}

\newc{\atmos}{\Delta m^2_{atm}}

\newc{\mnu}{m_{\nu}}

\newc{\lam}{\lambda}

\newc{\mM}{\mathcal{M}}

\newc{\lag}{\mathcal{L}}

\def\lsim{\raise0.3ex\hbox{$\;<$\kern-0.75em\raise-1.1ex\hbox{$\sim\;$}}}

\def\gsim{\raise0.3ex\hbox{$\;>$\kern-0.75em\raise-1.1ex\hbox{$\sim\;$}}}

\begin{document}

\begin{flushright}DO-TH-10/18\end{flushright}

\title{Lepton number, black hole entropy and $10^{32}$ copies of the Standard
Model
\footnote{Talk presented by Heinrich P\"as.} }

\author{Sergey Kovalenko$^1$, Heinrich P\"as$^2$ and Ivan Schmidt $^3$}

\address{$^1$,$^3$ Departamento de F\'{i}sica, 
Universidad T\'{e}cnica Federico Santa Mar\'{i}a,
Casilla 110-V, Valpara\'{i}so,
Chile and Center of Subatomic Physics,
 Valpara\'{i}so, Chile}
\address{$^2$ Fakult\"at Physik, Technische Universi\"at Dortmund, 44221 Dortmund, Germany}

\ead{$^1$ sergey.kovalenko@usm.cl}
\ead{$^2$ heinrich.paes@tu-dortmund.de}
\ead{$^3$ ivan.schmidt@usm.cl}

\begin{abstract}
Lepton number violating processes are a typical problem in theories
with a low quantum gravity scale. In this paper
we examine lepton number violation (LNV) in theories with a saturated black 
hole bound on a large number of species. Such theories have been 
advocated recently as a possible solution
to the hierarchy problem and an explanation of the smallness of neutrino 
masses. Naively one would expect black holes to 
introduce TeV scale LNV operators, thus 
generating unacceptably large rates of LNV processes. 
We show, however, that this does not happen in this scenario 
due to a complicated compensation mechanism between contributions
of different Majorana neutrino states to these processes.
As a result rates of LNV processes 
are extremely small and far beyond experimental reach, at least for the
left-handed neutrino states.
\end{abstract}

Recently the existence of
a large number of copies of Standard Model particles has
been proposed as a possibility to lower the Planck scale 
and solve the electroweak hierarchy problem \cite{Dvali:2007hz,Dvali:2007wp}.
In the following
we briefly review the main argument of this approach, closely following
\cite{Dvali:2007wp}: 
Let us assume there exist $N$ copies of the Standard Model, each carrying a 
separately conserved charge,
where $N$
is a large number $\sim 10^{32}$. It is in principle possible to prepare
a black hole containing one particle of each species. Now, due to charge 
conservation, the contained charges have to be revealed
in the evaporation process via Hawking radiation. On the other hand,
a particle of mass $\Lambda$ can only be emitted if the Hawking temperature
is large enough,
\beq
T_H \simeq \frac{M_P^2}{M_{BH}} \gsim \Lambda.
\label{th}
\eeq
Moreover, energy conservation bounds the maximal number of particles
with mass $\Lambda$
emitted due to the evaporation process of a black hole of mass $M_{BH}$
to be
\beq
n_{\rm max}=\frac{M_{BH}}{\Lambda}.
\eeq
Assuming that the number of states $N$ saturates the black hole bound we
obtain with (\ref{th})
\beq
N=n_{\rm max}=\frac{M_P^2}{\Lambda^2}.
\eeq
Finally, since semiclassically the lifetime of a blck hole is given by
\beq
\tau_{BH}=\frac{1}{N}\int \frac{1}{T_H^2} d M_{BH} \sim \Lambda^{-1}
\eeq
we conclude that a black hole of size $\Lambda$ has a life time
of order $\Lambda^{-1}$ implying that $\Lambda$ is the scale where the
semi-classical treatment breaks down and quantum gravity sets in, namely
the true Planck scale,
\beq
\Lambda \simeq \frac{M_P}{\sqrt{N}},
\eeq
where $M_P$ has to be interpreted as the effective Planck scale, which 
implies $\Lambda \sim {\cal O}(TeV)$ for
$N\simeq 10^{32}$ and thus solves the hierarchy problem. 

Finally, in \cite{Dvali:2009ne} this scenario has been advocated also
as a mechanism for generating small neutrino masses, providing an attractive 
alternative for seesaw, extra dimensional
and other known mechanisms. 
It is assumed that there exists one SM singlet right-handed neutrino
$\nu_{Rj}$ per SM$_{j}$ copy, so that $j=1,...,N$.  The mechanism relies on the fact 
that the right-handed neutrinos, being SM singlets, couple to all the SM copies ``democratically''. 
This SM singlet democracy combined with the requirement of unitarity of the theory 
implies a $1/\sqrt{N}$ suppression of the corresponding Yukawa couplings to the left-handed 
neutrinos  $\nu_{Lj}$ and thus a suppression of the Dirac type neutrino masses.
The minimalistic approach to the problem of smallness of neutrino mass suggests that 
in this scenario possible $B-L$ violating Majorana masses of $\nu_{Rj}$ are unnecessary and the lepton number is conserved.


However, the assumption of lepton number conservation appears rather ad hoc,
as there is no fundamental reason to forbid Majorana masses for 
the right-handed neutrinos and lepton number conservation appears  as accidental symmetry. 
Quantum gravity breaks global symmetries, and then conserved lepton number requires a gauged $B-L$ symmetry 
$U_{1 (B-L)}$. The latter should be spontaneously broken otherwise there must exist the corresponding massless 
gauge boson stringently constrained by phenomenology.  On the other hand lepton number violation might be 
helpful for successful baryogenesis. 

In the following we analyze some consequences of the $N$-copies SM without  lepton number conservation
. Interestingly, even in this case, Majorana masses turn out to be suppressed by an exactly analogous reason as the Dirac masses. 
We assume a 
\begin{eqnarray}\label{symmetry}
\prod_{i}\left(SU_{3c}\times SU_{2W}\times U_{y}\right)_{i}\times U_{1 (B-L)}\times Z_{N}
\end{eqnarray}
gauge symmetry of the $N$-copies SM. 
It includes a common anomaly free gauge  factor 
$U_{1 (B-L)}$.  We introduce this gauge symmetry in order to prevent 
the appearance of phenomenologically dangerous 
LNV operators induced by the TeV black holes.  
An additional permutation symmetry $Z_{N}$ acting in the space of the 
SM$_{i}$ species is also imposed   \cite{Dvali:2009ne} in the $N$-copy 
SM scenario. 
To be unaffected by black holes, this discrete symmetry 
should be considered as a gauged symmetry in the sense of being a discrete 
subgroup of some continuous gauge group.  
The Lagrangian terms relevant for our discussion are the following:
\begin{eqnarray}\label{Lag}
{\cal L}_{\nu H S} = \lambda_{ij} \overline{\nu_{Rj}}\left(L H \right)_{i} + \beta_{ij} \overline{\nu_{Ri}^{c}} \nu_{Rj} S+ \kappa_{i} (H^{\dagger} H)_{i} S^{\dagger} S. 
\end{eqnarray}
The model involves $N$ right-handed SM singlet neutrinos $\nu_{R i}$ and one SM singlet complex 
scalar field $S$ having the $B-L$-charge equal to +2. Then the trilinear $HHS$ couplings are forbidden in (\ref{Lag}). 
The $U_{1 (B-L)}$ is spontaneously broken by a vacuum expectation value $\langle S\rangle$ resulting 
in a Majorana mass term from the second term in Eq. (\ref{Lag}). We assume that the scale of 
$B-L$ breaking lies below the gravity cutoff $\Lambda$. The Dirac mass term considered in ref. \cite{Dvali:2009ne} 
originates from the first term after the electroweak symmetry breaking.  

Let us consider the $N\times N$ Yuakawa coupling
matrix $\lambda_{ij}$  of Dirac type following 
ref. \cite{Dvali:2009ne}.
As the $\nu_{Ri}$ fields are not charged under the SM symmetry they cannot
be assigned to a single SM-copy, besides respecting the same transformation
properties under a permutation symmetry acting on the space of species.
This permutation symmetry constrains the Yukawa coupling matrix to the form 
\beq
 \label{lambda}
\lambda_{ij}=\left(
\begin{array}{cccc}
a & b & b & ..\\
b & a & b & ..\\
b & b & a & ..\\
.. & .. & .. & ..\\ 
\end{array}
\right).
\eeq
This matrix combined in the first term in (\ref{Lag}) with the SM Higgs expectation 
value $\langle H_{i}\rangle = \langle H\rangle $ results in the
Dirac neutrino mass matrix 
\begin{eqnarray}
\label{M-D}
m^{D}_{ij}=\lambda_{ij}\langle H\rangle.
\end{eqnarray}
Following ref. \cite{Dvali:2009ne} we assume that the electroweak symmetry breaking leaves the permutation symmetry unbroken. This implies that the  VEVs of all the Higgs species are equal to the same value $\langle H\rangle$.  
A key point guaranteeing 
the smallness of neutrino mass matrix entries is the smallness of the Yukawa coupling matrix (\ref{lambda}) which follows from the requirement of unitarity of the theory. In fact, let us consider the right-handed neutrino inclusive production in the scattering of the SM particles as shown in Fig. 1(a).   
At high energies the rate of this process grows  like
\beq
\label{BD-N}
\Gamma \simeq N b^2 E,
\eeq
as follows from dimensional analysis. Here we assumed $a\sim b$ which is suggested by an observation that these two quantities are of the same nature and have no fundamental reason to be very different in magnitude \cite{Dvali:2009ne}.
Unitarity below the gravity cutoff is preserved only for 
\beq\label{b-N}
b\lsim \frac{1}{\sqrt{N}}.
\eeq
Thus the neutrino mass matrix (\ref{M-D}) results  in $N-1$ Dirac neutrinos with tiny masses 
$m_D \simeq  \langle H\rangle/\sqrt{N} \lsim {\cal O}(eV)$ \cite{Dvali:2009ne} 
which fulfill the experimental bounds constraining them to the sub-eV scale. One neutrino state in this framework is very heavy with the mass of the order 
$M^{D}\simeq \sqrt{N} \langle H\rangle$ which is comparable with the Plank scale.  
\begin{figure}[!t]
\centerline{
    \includegraphics[scale=0.6]{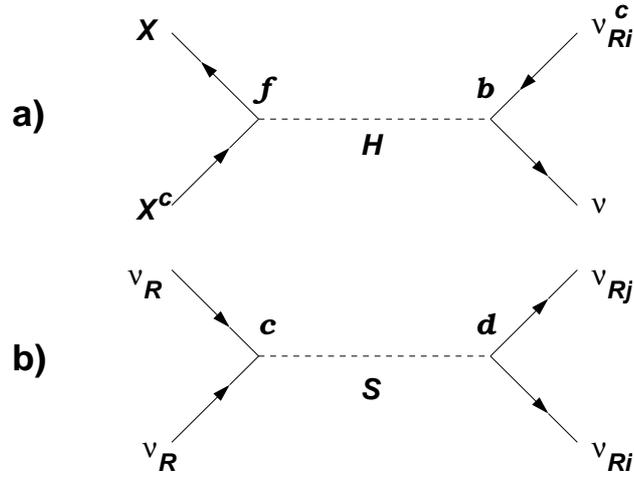}}
\caption{Diagrams relevant for Unitarity constraints on Neutrino Yukawa couplings:
(a) to the Higgs doublet $H_{i}$ (Lepton number conserving); (b) to a Higgs Singlet 
(Lepton number violating). $X_i$ denotes some of the SM fields coupled to 
$H_{i}$ with a strength $f$.    
 \label{Large-N}}
\end{figure}

\begin{figure}[!t]
\centerline{
    \includegraphics[scale=0.6]{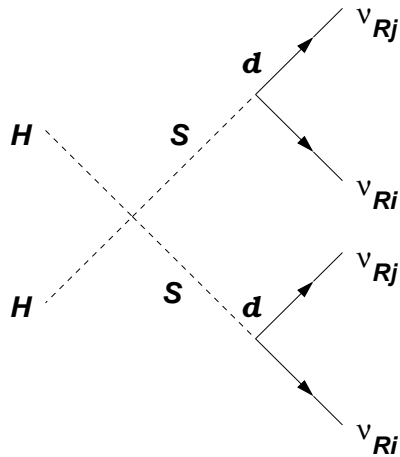}}
\caption{The diagram leading to the strongest Unitarity constraint on the Majorana neutrino mass term.
 \label{scalar}}
\end{figure}

\begin{figure}[!t]
\smallskip
\centerline{
    \includegraphics[scale=0.6]{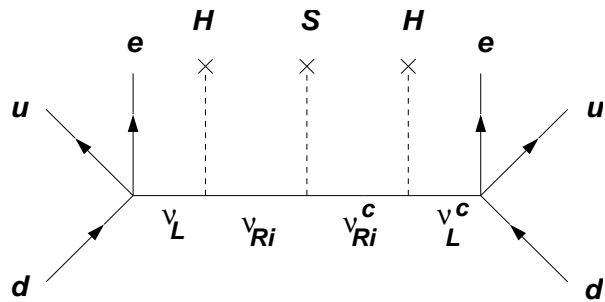}}
\caption{Diagram for neutrinoless double beta decay in the presence
of $N$ copies of the SM particle content. 
 \label{0vbb-N}}
\end{figure}

Now let us turn to the second term of Eq. (\ref{Lag}) which after the 
$B-L$-symmetry breaking leads to the Majorana mass matrix of the right-handed neutrinos
\begin{eqnarray}
\label{M-M}
m^{M}_{ij} = \beta_{ij} \langle S\rangle
\end{eqnarray}
The permutation symmetry constrains the Majorana type 
Yukawa coupling $N\times N$ matrix, 
just as the Dirac type Yukawa couplings had been constrained before,
to be of the form
\beq
\beta_{ij}=
\left(
\begin{array}{cccc}
c & d & d & ..\\
d & c & d & ..\\
d & d & c & ..\\
.. & .. & .. & ..\\ 
\end{array}
\right).
\eeq
The similarity arguments used above to justify $a\sim b$ in Eqs. (\ref{lambda}), 
(\ref{BD-N}) can equally be applied to motivate $d\sim c$. 
Now, when considering the scattering process of right-handed neutrinos,
both final states can be any of the $N$ copies as in Fig. 1(b) and,  thus, the inclusive rate
grows like  
\beq
\Gamma \simeq N^2 c^2 d^2 E,
\eeq
which preserves unitarity below the gravity cutoff only for 
\beq \label{c-N}
c \sim d \lsim \frac{1}{\sqrt{N}}.
\eeq
An even  more stringent bound originates from the diagram of Higgs doublet scattering
in Fig.~\ref{scalar},
\beq
\Gamma \simeq N^4 \kappa^{2}d^4 E,
\eeq
The scalar quartic couplings $\kappa_{i}$ should not be very small since there is no symmetry protecting its smallness. 
 Thus, an assumption  $\kappa_{i} \sim 1$ could be used for rough estimations. Then we have
\beq\label{cd-lim}
c\sim d \lsim \frac{1}{N}.
\eeq
This limit remains unaffected by insertion of additional $S$-branches 
in the diagram in Fig. 2.
As a consequence the neutrino Majorana mass matrix entries 
(\ref{M-M}) are even more strongly suppressed than the Dirac masses
discussed above. 

In order to discuss the phenomenology of LNV processes we now analyze
the
mass spectrum of the present scenario. 
The mass matrix written in the basis of the 2$N$ fields
${\cal N}= \{\nu_{Li}, \nu_{Ri}\}$ has the following form
\begin{eqnarray}\label{MassMatr-1}
{\cal M}^{\nu}=
\left(
\begin{array}{cccc}
0&m^{D}\\
m^{D} &m^{M}\\ 
\end{array}
\right).
\end{eqnarray}
where $m^{D}$ and $m^{M}$ are $N\times  N$ submatrices 
given by Eqs. (\ref{M-D})-(\ref{M-M}).  The set of $2N$ mass eigenstates 
$\nu_{i}=U_{ij}{\cal N}_{j}$ of this symmetric matrix split in 
two groups of $(N-1)$ degenerate states $\nu^{+}$ and $\nu^{-}$ and 
another two heavy states $N^{\pm}$. 
The details of the diagonalization procedure will be given
elsewhere \cite{kov}.

For consistency with neutrino phenomenology one needs one 
neutrino at the sub-eV scale, say, 
$m_{\nu} \sim 10^{-2}$eV. Then, 
the states $N^{\pm}$ are pushed in mass towards the Planck scale and, 
therefore,  their phenomenological impact is negligible. 
The light Majorana states $\nu^{\pm}$ having very small mass splitting  $\delta_{m}\sim 1/\sqrt{N}$ form a quasi Dirac state 
with the mass $m_{\nu}$. Thus they are expected to induce lepton number violating processes at 
rates $\sim 1/N$. 
Moreover due to the structure of the mass matrix  (\ref{MassMatr-1})  with zero submatrix in the upper-left corner 
LNV processes are even more strongly suppressed. 
Thus the dominating contributions to neutrinoless double beta decay are
\begin{eqnarray}\label{m3}
\langle m_{\nu}^{3}\rangle &=& (a-b)^{2}(c-d)\langle H\rangle^{2}\langle S\rangle\sim N^{-2}, \\
\nonumber 
\left\langle \frac{1}{M_{N}}\right\rangle &\approx& \frac{\langle H\rangle}{\langle S\rangle^{2}}\frac{d}{b^{2} N^{2}} \sim  N^{-2}.
\end{eqnarray} 
Therefore, the amplitude for neutrinoless double beta decay
and other, related processes in the studied framework is extremely small.  

In this paper we have adressed a problem arising in any scenario
with a low quantum gravity scale: do LNV operators induced by TeV scale
black holes invalidate the model?
For the case of the $N=10^{32}$-copies SM we have shown that this
consequence is avoided due to a non-trivial cancelation mechanism. This
property should be considered as an important benefit of the model.

Nevertheless, the presence of a large number of right-handed Majorana
states can have interesting phenomenological consequences.
For example, a very naive estimate of the right-handed neutrino decay diagrams
on tree and one-loop level, which give rise to the baryon asymmetry in
leptogenesis, scale as $(\sqrt{N})^2$ from the Yukawa coupling with
$N$ $\nu_{Ri}$ copies contributing potentially to the decay, and the $\nu_{Ri}$
propagator in the loop diagram. So the process may be relevant despite LNV
signals being suppressed for the left-handed neutrino states.
A similar line of reasoning could apply to single $\nu_{Ri}$ production
at the LHC. 

We thus conclude that the $N=10^{32}$-copies SM is safe from LNV 
in the SM sector,
and leave the potentially interesting phenomenology of $\nu_{Ri}$
production and decay for further study.

\section*{Acknowledgments}
This work has been partially supported by the DFG grant PA-803/5-1
and by the PBCT project ACT-028 ``Center of Subatomic Physics" and  
CONICYT, Programa de Financiamiento Basal para Centros 
Cient\'{i}ficos y Tecnol\'ogicos de Excelencia.
HP thanks the  Universidad T\'{e}cnica Federico Santa Mar\'{i}a
for hospitality offered while part of this work was carried out.

\bibliographystyle{iopart-num}

\begin{thebibliography}{17}


\bibitem{Dvali:2007hz}  G.~Dvali,
 arXiv:0706.2050 [hep-th].

\bibitem{Dvali:2007wp}  G.~Dvali and M.~Redi,
 Phys.\ Rev.\  D {\bf 77}, 045027 (2008)
[arXiv:0710.4344 [hep-th]].

\bibitem{Dvali:2008ec}
  G.~Dvali and C.~Gomez,
  Phys.\ Lett.\  B {\bf 674}, 303 (2009)
  [arXiv:0812.1940 [hep-th]].

\bibitem{Dvali:2009ne}  G.~Dvali and M.~Redi,
 Phys.\ Rev.\  D {\bf 80}, 055001 (2009),
  [arXiv:0905.1709 [hep-ph]].

\bibitem{kov}
S. Kovalenko, H. P\"as, I. Schmidt, to be published




\end{thebibliography}

\end{document}